\documentclass[a4paper,fleqn]{cas-dc}

\usepackage[numbers]{natbib}

\usepackage{amssymb}

\usepackage{graphicx}
\usepackage{amsmath}
\usepackage{subfigure}
\usepackage{bbold}
\usepackage{yfonts}
\usepackage{placeins}
\usepackage{bm} 
\usepackage{nicefrac}
\usepackage{slashed}

\def\be{\begin{equation}}
	\def\ee{\end{equation}}
\newcommand{\bel}[1]{\begin{eqnarray}\label{#1}}
	\newcommand{\eel}{\end{eqnarray}}
\def\barr{\begin{array}}
	\def\earr{\end{array}}
\def\beq{\begin{eqnarray}}
	\def\eeq{\end{eqnarray}}
\def\bfig{\begin{figure}}
	\def\efig{\end{figure}}

\newcommand{\rf}[1]{Eq.~(\ref{#1})}
\newcommand{\rfm}[1]{Eqs.~(\ref{#1})}

\newcommand{\rfn}[1]{(\ref{#1})}

\def\g{\gamma}

\def\c{\chi}
 
\newcommand{\bea}{\begin{eqnarray}}
\newcommand{\eea}{\end{eqnarray}}

\newcommand{\av}{{\boldsymbol a}} 
\newcommand{\bv}{{\boldsymbol b}} 

\newcommand{\xv}{{\boldsymbol x}}

\newcommand{\pv}{{\boldsymbol p}}

\def\S0iU{{\Sigma}^{0i}}

\def\n0{n_{(0)}}
\def\e0{\varepsilon_{(0)}}
\def\P0{P_{(0)}}

\def\dabxy{\delta_{ab} \delta^{(3)}({\boldsymbol x}-{\boldsymbol y})}

\def\psix{\psi(t,{\boldsymbol x})}
\def\psixdag{\psi^\dagger(t,{\boldsymbol x})}
\def\bpsix{\bar{\psi}(t,{\boldsymbol x})}

\def\psiax{\psi_a(t,{\boldsymbol x})}

\def\psibx{\psi_b(t,{\boldsymbol x})}
\def\psiby{\psi_b(t,{\boldsymbol y})}
\def\psidagax{\psi^\dagger_a(t,{\boldsymbol x})}
\def\psidagby{\psi^\dagger_b(t,{\boldsymbol y})}

\newcommand{\ps}{\psi}
\newcommand{\psbar}{\bar{\psi}}
\newcommand{\olra}{\overleftrightarrow}

\newcommand{\dx}{\!d^3x\, }

\begin{document}
\let\WriteBookmarks\relax
\def\floatpagepagefraction{1}
\def\textpagefraction{.001}
\shorttitle{Pseudogauge freedom and the SO(3) algebra of spin operators}
\shortauthors{ }

\title [mode = title]{Pseudogauge freedom and the SO(3) algebra of spin operators}  




\author[1]{Sourav Dey}[orcid=]
\ead{sourav.dey@niser.ac.in}

\address[1]{School of Physical Sciences, National Institute of Science Education and Research, An OCC of Homi Bhabha National Institute, Jatni-752050, India}

\author[2]{Wojciech Florkowski}[orcid=0000-0002-9215-0238]
\ead{wojciech.florkowski@uj.edu.pl}

\address[2]{Institute of Theoretical Physics, Jagiellonian University, ul. St. \L ojasiewicza 11, PL-30-348 Krakow, Poland}

\author[1]{Amaresh Jaiswal}[orcid=0000-0001-5692-9167]
\ead{a.jaiswal@niser.ac.in}

\author[3]{Radoslaw Ryblewski}[orcid=0000-0003-3094-7863]
\ead{radoslaw.ryblewski@ifj.edu.pl}

\address[3]{Institute of Nuclear Physics Polish Academy of Sciences, PL-31-342 Krakow, Poland}



\begin{abstract}
The energy-momentum and spin tensors for a given theory can be replaced by alternative expressions that obey the same conservation laws for the energy, linear momentum, as well as angular momentum but, however, differ by the local redistribution of such quantities (with global energy, linear momentum, and angular momentum remaining unchanged). This arbitrariness is described in recent literature as the pseudogauge freedom or symmetry. In this letter, we analyze several pseudogauges used to formulate the relativistic hydrodynamics of particles with spin $\nicefrac{1}{2}$ and conclude that the canonical version of the spin tensor has an advantage over other forms as only the canonical definition defines the spin operators that fulfill the SO(3) algebra of angular momentum. This result sheds new light on the results encountered in recent papers demonstrating pseudogauge dependence of various physical quantities. It indicates that for spin-polarization observables, the canonical version is fundamentally better suited for building a connection between theory and experiment.
\end{abstract}


\maketitle
\section{Introduction}

Formulation of relativistic hydrodynamics with angular momentum conservation, termed relativistic spin hydrodynamics, has recently been the subject of intense investigation; see Refs.~\cite{Florkowski:2018fap, Speranza:2020ilk, Bhadury:2021oat} for recent review. In these formulations, spin degrees of freedom are incorporated with the help of a new hydrodynamic variable known as the spin tensor. It is well known, however, that the energy-momentum and spin tensors for a given theory can be replaced by different expressions that obey the same conservation laws for energy, linear momentum, and angular momentum. The new expressions differ by local redistribution of such quantities, but the total energy, along with the total linear and angular momenta remain unchanged~\cite{Leader:2013jra}. 

This arbitrariness is known in recent literature as the \textit{pseudogauge freedom} or symmetry and is defined by the two transformations~\cite{Hehl:1976vr}
\begin{eqnarray}
T^{\prime \mu \nu}&\!\!=\!\!& T^{\mu\nu} + \frac{1}  {2} \partial_\lambda \left( 
\Phi^{\lambda, \mu \nu} 
+\Phi^{\nu, \mu \lambda}
+\Phi^{\mu, \nu \lambda} \right), \nonumber \\
S^{\prime \lambda, \mu \nu}&\!\!=\!\!&
S^{\lambda, \mu \nu} - \Phi^{\lambda, \mu \nu} + \partial_\rho Z^{\mu\nu, \lambda \rho}. \label{eq:PG}
\end{eqnarray}
Here $T^{\mu\nu}$ ($S^{\lambda, \mu \nu}$) and $T^{\prime \mu \nu}$ ($S^{\prime \lambda, \mu \nu}$) are the energy-momentum (spin) tensors before and after the pseudogauge transformation, while the quantities $\Phi^{\lambda, \mu \nu}$ and $Z^{\mu\nu, \lambda \rho}$ are tensors known as superpotentials. They have the following symmetries with respect to the exchange of indices
\begin{eqnarray}
\Phi^{\lambda, \mu \nu} &=& -\Phi^{\lambda, \nu \mu}, \nonumber \\
Z^{\mu\nu, \lambda \rho} &=& - Z^{\nu\mu, \lambda \rho} \,\,\,=\,\,\, -Z^{\mu\nu, \rho \lambda}. \label{eq:PGsymmetries}
\end{eqnarray}
All quantities in \rfm{eq:PG} are constructed from the field operators. The most prominent example of the pseudogauge symmetry is the Belinfante transformation~\cite{BELINFANTE1939887,Belinfante:1940,rosenfeld1940tenseur}, where \rf{eq:PG} is used with $Z^{\mu\nu, \lambda \rho}=0$ and $\Phi^{\lambda, \mu \nu} = S^{\lambda, \mu \nu}$. This leads to a symmetric energy-momentum tensor and a vanishing spin tensor.

In this letter, we analyze three different versions of the energy-momentum and spin tensors that have been recently used to formulate relativistic hydrodynamics of particles with spin $\nicefrac{1}{2}$: the standard canonical versions of $T^{\mu\nu}$ and $S^{\lambda, \mu \nu}$~\cite{Itzykson:1980rh}, the versions of de Groot, van Leuween and van Weert~\cite{DeGroot:1980dk}, denoted as $T^{\mu\nu}_{\rm GLW}$ and $S^{\lambda, \mu \nu}_{\rm GLW}$, and the versions of Hilgevoord  and Wouthuysen~\cite{HILGEVOORD19631,HILGEVOORD19651002}, denoted as $T^{\mu\nu}_{\rm HW}$ and $S^{\lambda, \mu \nu}_{\rm HW}$. For each of these three versions (pseudogauges), we check if the spin operators can be considered as ``good'' angular momentum operators. On general grounds, one expects that a set of angular momentum operators representing total ($\textbf{J}$), orbital ($\textbf{L}$), and spin ($\textbf{S}$) parts of angular momentum satisfies the fundamental SO(3) algebra for equal-time commutation relations:
\begin{eqnarray}
\left[J^i(t),J^j(t)\right] &=& i\varepsilon^{ijk}J^k(t), \label{eq:com_rel_J}\\
\left[L^i(t),L^j(t)\right] &=& i\varepsilon^{ijk}L^k(t), \label{eq:com_rel_L}\\ 
\left[S^i(t),S^j(t)\right] &=& i\varepsilon^{ijk}S^k(t). \label{eq:com_rel_S}
\end{eqnarray}
A failure to correctly reproduce the commutation relations \rfn{eq:com_rel_J}--\rfn{eq:com_rel_S} in a given theory may lead to fallacious or at least misleading conclusions~\cite{LEADER201923}.

As the pseudogauge transformations do not change the total conserved ``charges'', their commutation relations remain unchanged. In particular, this means that the commutation relation \rfn{eq:com_rel_J} is pseudogauge invariant. On the other hand, nothing protects the commutation relations \rfn{eq:com_rel_L} and \rfn{eq:com_rel_S} describing two separate contributions to the total angular momentum. It is known~\cite{Leader:2013jra, LEADER201923} that Eqs.~\rfn{eq:com_rel_L} and \rfn{eq:com_rel_S} hold in the case of the canonical spin tensor and, trivially, in the Belinfante case (where the spin tensor vanishes). On the other hand, it is not obvious if the SO(3) algebra is fulfilled by the GLW and HW spin operators. 

To be more specific, we define the total spin operator as the integral
\begin{equation}
\frac{1}{2} S^k(t) = \frac{1}{2} \varepsilon^{kij} \int\dx S^{0, ij}\left(t, \xv\right) \label{eq:Sk_} = \frac{1}{2} \varepsilon^{kij} S^{ij}(t)
\end{equation}
and verify the property~\footnote{We keep $\nicefrac{1}{2}$ factored out so that \rf{eq:SO3} looks similar to the commutation relation for the Pauli sigma matrices. The symbols without subscripts denote the canonical versions.}
\begin{equation}
\left[\frac{1}{2} S^i(t),\frac{1}{2} S^j(t) \right] =
i \varepsilon^{ijk} \frac{1}{2} S^k(t). \label{eq:SO3}
\end{equation}
We note that the angular momentum operators are functions of the field operators of the theory. In order to check whether the commutation relation \rfn{eq:SO3} holds, one must know the fundamental commutation relations between fields and their conjugate momenta. 

In this work, we take into account a gas of relativistic fermions with spin $\nicefrac{1}{2}$ described by the Dirac equation. Although the considered system is quite simple, it is commonly considered a starting point for the construction of relativistic hydrodynamics of spin-polarized media~\cite{Florkowski:2018fap}. It also allows for an exact study of the pseudogauge dependence of the commutation relation \rfn{eq:SO3}. Out of the three cases studied herein (canonical, GLW, and HW), only the canonical version satisfies \rf{eq:SO3}. Thus, we conclude that the canonical version of the spin tensor has a very specific advantage over other forms used in the literature. In particular, for the spin-polarization observables, the canonical version is better suited for building a connection between theory and experiment.

\medskip
Throughout the paper we use the Dirac representation for Dirac matrices as well as the convention $\varepsilon^{0123}=+1$. The metric tensor has the signature $(+,-,-,-)$. Three-vectors are denoted by the bold font. The scalar products for both three- and four-vectors are denoted by a dot, i.e., $a \cdot b = a^0 b^0 - \av \cdot \bv$.
\section{Canonical, GLW, and HW versions of the energy-momentum and spin tensors}

The standard application of Noether's Theorem to the Dirac Lagrangian density
\begin{equation}
\label{ld}
\mathcal{L}_D(x)=\frac {i} 2 \,  \psbar (x)\gamma^\mu\olra{\partial}_\mu\ps (x) - m\psbar (x)\ps (x)
\end{equation}
yields the conserved \textit{canonical} energy-momentum tensor
\begin{equation}
T^{\mu\nu} = \frac{i}{2}\psbar\gamma^\mu \olra{\partial}^\nu\ps - g^{\mu\nu}\mathcal{L}_D 
\label{tcan}
\end{equation}
and the conserved total angular momentum operator
\begin{equation}
\label{js}
J^{\lambda, \mu\nu}  = L^{\lambda,\mu\nu} + S^{\lambda,\mu\nu},
  \end{equation}
where $L^{\lambda,\mu\nu} = x^\mu T^{\lambda\nu}-x^\nu T^{\lambda\mu}$ is the orbital angular momentum, and $S^{\lambda,\mu\nu}$ is the spin tensor defined by the expression~\cite{Itzykson:1980rh}
\begin{equation}
S^{\lambda, \mu \nu}=\frac{1}{4} \bar{\psi}\left\{\gamma^\lambda, \sigma^{\mu \nu}\right\} \psi=-\frac{1}{2} \varepsilon^{\lambda \mu \nu \alpha} \bar{\psi} \gamma_\alpha \gamma_5 \psi.
\end{equation}
Here $\sigma^{\mu \nu}\equiv\frac{i}{2}\left[ \gamma^\mu, \gamma^\nu\right]$ and $\overleftrightarrow{\partial}_\mu  \equiv \overrightarrow{\partial}_\mu -\overleftarrow{\partial}_\mu$.

We note that the conservation of total angular momentum expressed by the equation $\partial_\lambda J^{\lambda, \mu\nu} = 0$ is equivalent to the formula
\begin{equation}
\partial_\lambda S^{\lambda, \mu \nu}
= T^{\nu\mu} - T^{\mu\nu}.
\label{eq:spinnoncon}
\end{equation}
Thus, for an asymmetric energy-momentum tensor, which is the case for the canonical version, the spin tensor is not conserved (even for free particles). This property was interpreted as a deficiency of the canonical forms of the energy-momentum and spin tensors since one naively expects that the spin part of the angular momentum of non-interacting particles should be conserved. As a possible solution to this problem, de Groot, van Leeuwen, and van Weert proposed~\cite{DeGroot:1980dk} to switch from the canonical forms to alternative forms with the help of the pseudogauge transformation \rfn{eq:PG} with the superpotentials~\cite{Speranza:2020ilk} 
 \begin{eqnarray}
\Phi^{\lambda,\mu\nu}_{\rm GLW}&=&\frac{i}{4m}\bar{\psi}\left(\sigma^{\lambda\mu} \overleftrightarrow{\partial}^\nu-\sigma^{\lambda\nu}\overleftrightarrow{\partial}^\mu\right)\psi,
\label{eq:PhiGLW} \\
Z_{\rm GLW}^{\mu\nu,\lambda\rho}&=&0. 
\label{eq:ZGLW}
\end{eqnarray}

 An alternative version of the pseudogauge transformation that also leads to a conserved spin tensor was proposed by Hilgevoord and Wouthuysen~\cite{HILGEVOORD19631,HILGEVOORD19651002}. In this case, the superpotentials read~\cite{Speranza:2020ilk}
 \begin{eqnarray}
\Phi^{\lambda,\mu\nu}_{\rm HW}&=&
\frac{i}{4m}\psbar \left( \sigma^{\lambda\mu}\olra{\partial}^\nu
-\sigma^{\lambda\nu}\olra{\partial}^\mu
\right)\ps \nonumber \\
& & - \frac{i}{4m}\psbar 
\left( 
g^{\lambda \mu } \sigma^{\nu\rho}
- g^{\lambda \nu } \sigma^{\mu\rho}
\right) \olra{\partial}_\rho \psi, \label{eq:PhiHW}
\\
Z^{\mu\nu\lambda\rho}_{\rm HW} &=& -\frac{1}{8m}\psbar\left(\sigma^{\mu\nu}\sigma^{\lambda\rho}+\sigma^{\lambda\rho}\sigma^{\mu\nu}\right)\ps , \label{eq:ZHW}
\end{eqnarray}
Note that the first line in \rf{eq:PhiHW} agrees with the definition of $\Phi_{\rm GLW}$. 

\section{Spin tensors and the SO(3) algebra}

Following \rf{eq:Sk_} we define the total spin operator in the GLW pseudogauge as
\begin{equation}
 \frac{1}{2} S^k_{\rm GLW}(t) = \frac{1}{2} \varepsilon^{kij} \int\dx \left(S^{0, ij}(t, \xv) - \Phi^{0,ij}_{\rm GLW}(t, \xv) \right)
\label{eq:SkGLW}
\end{equation}
and, correspondingly, the total spin operator in the HW pseudogauge
\begin{eqnarray}
\frac{1}{2} S^k_{\rm HW}(t) \!\!\!&=&\!\!\! \frac{1}{2} \varepsilon^{kij} \int\dx \left(S^{0, ij}(t, \xv) - \Phi^{0,ij}_{\rm HW}(t, \xv) \right) 
\nonumber \\
&&\hspace{-0.4cm} +  \frac{1}{2} \varepsilon^{kij} \int\dx \partial_\rho Z^{ij,0 \rho}_{\rm HW}(t, \xv).
\label{eq:SkHW}
\end{eqnarray}
%
Using general symmetry properties of the superpotentials listed in \rf{eq:PGsymmetries} and assuming that the whole system under consideration is localized (boundary terms in the integration by parts can be neglected) we find that the term in the second line in \rf{eq:PhiHW}, as well as the expression \rfn{eq:ZHW}, do not contribute to $S^k_{\rm HW}(t)$, hence, we obtain
\begin{equation}
  S^k_{\rm GLW}(t) = S^k_{\rm HW}(t).
\end{equation}
Thus, our task remains to check the SO(3) algebra only for the GLW case, namely, we are going to verify if
\begin{equation}
\left[\frac{1}{2} S^i_{\rm GLW}(t),\frac{1}{2} S^j_{\rm GLW}(t) \right] \overset{?}{=}
i \varepsilon^{ijk} \frac{1}{2} S^k_{\rm GLW}(t). \label{eq:SO3GLW}
\end{equation}
%

\subsection{Canonical formulation}

Let us first show that \rf{eq:SO3} holds for the canonical case, where
\begin{eqnarray}
 S^{ij}= \frac{1}{2} \varepsilon^{0ijk} \int\dx \bpsix \,\gamma^k \gamma_5 \psix,
\end{eqnarray}
which leads us to the definition
\begin{eqnarray}
S^k(t)=\int\dx \psidagax \,\Sigma_{a b}^k \psibx.
\label{eq:Sk}
\end{eqnarray}
Here $a$ and $b$ are spinor indices and~\cite{Itzykson:1980rh}
\begin{equation}
\Sigma^k = \gamma_5 \gamma_0 \gamma^k = 
\left(
\begin{matrix}
\sigma^k & 0 \\
0 & \sigma^k 
\end{matrix}
\right). \label{eq:Sigmak}
\end{equation}
To prove \rf{eq:SO3} we use the formula \rfn{eq:Sk} in \rf{eq:SO3} and apply twice the equal-time commutation relations for the Dirac field~\cite{Coleman:2018mew}
\begin{eqnarray}
\{ \psiax, \psidagby \} &=& \dabxy, \label{eq:anticom} \\
\{ \psiax, \psiby \} &=& \{\psidagax, \psidagby \} \,=\, 0. \nonumber
\end{eqnarray}
This leads us to the expression
\begin{equation}
\left[S^i(t),S^j(t)\right] = \int\dx \psixdag 
\left[ \Sigma^i, \Sigma^j \right] \psix.
\end{equation}
Since $\left[ \Sigma^i, \Sigma^j \right] = 2 i \varepsilon^{ijk}\Sigma^k $, we immediately reproduce \rf{eq:SO3}. Thus, the spin operators defined for the canonical case are indeed the angular momentum operators.
 
\subsection{GLW formulation}

In the case of the GLW decomposition, following Eqs.~\rfn{eq:PhiGLW} and \rfn{eq:ZGLW}, we define
\begin{equation}
S^{ij}_{\rm GLW}(t) = S^{ij}(t) -\Phi^{ij}_{\rm GLW}(t)
\end{equation}
and
\begin{eqnarray}
 \Phi^m_{\rm GLW}(t) &=&  \varepsilon^{mij} \Phi^{ij}_{\rm GLW}(t)  \\
 \hspace{1cm} &=& \frac{i}{m} \varepsilon^{mij}
\int\dx\psixdag
{\tt S}^i \partial^j
 \psix. \nonumber
\end{eqnarray}
Here we have introduced the notation~\cite{Itzykson:1980rh}
\begin{equation}
{\tt S}^i = \gamma_0 \sigma^{0i} = i
\left(
\begin{matrix}
0 & \sigma^i \\
-\sigma^i & 0 
\end{matrix}
\right). \label{eq:calSi}
\end{equation}
The matrices $\Sigma$ and ${\tt S}$ have the following commutation relations
\begin{eqnarray}
\left[\Sigma^m, {\tt S}^k \right] &=& 2 i \varepsilon^{mkj} {\tt S}^j
\label{eq:ttSSig}, \\
\left[{\tt S}^i, {\tt S}^k \right] &=& 2 i \varepsilon^{ikr} \Sigma^r.
\label{eq:ttSttS}
\end{eqnarray}
This property can be checked by a direct calculation in which explicit expressions for the Dirac gamma matrices are used. The relation \rfn{eq:SO3GLW} holds if
\begin{eqnarray}
&& \hspace{-1.0cm} \left[S^m(t), \Phi_{\rm GLW}^n(t) \right]
- \left[S^n(t), \Phi_{\rm GLW}^m(t) \right] \nonumber \\
&&
- \left[\Phi_{\rm GLW}^m(t), \Phi_{\rm GLW}^n(t) \right]
= 2i \varepsilon^{mnk} \Phi_{\rm GLW}^k(t). \label{eq:SO3GLW1}
\end{eqnarray}
Here we used \rf{eq:SO3} for the canonical spin operators. Again, using the anticommutation rules for the field operators \rfn{eq:anticom} (after differentiating them with respect to the spatial coordinates $x^i$) we can show that
\begin{eqnarray}
&& \hspace{-1.0cm} \left[S^m(t), \Phi_{\rm GLW}^n(t) \right] 
= \nonumber  \\
&& 
\frac{i}{m} \varepsilon^{nkl}\!\!
\int\dx \psixdag \left[    
\Sigma^m, {\tt S}^k  \right] \partial_x^l \psix, \nonumber
\end{eqnarray}
where $\partial_x^l \equiv \partial/\partial x_l$. Using now the commutation relation \rfn{eq:ttSSig} we find that
\begin{eqnarray}
 \left[S^m(t), \Phi_{\rm GLW}^n(t) \right]
- \left[S^n(t), \Phi_{\rm GLW}^m(t) \right] 
= 2i \varepsilon^{mnk} \Phi_{\rm GLW}^k(t). \label{eq:SO3GLW2}
\end{eqnarray}
Consequently, the identity \rfn{eq:SO3GLW} holds if the commutator in the second line of \rf{eq:SO3GLW1} vanishes. To calculate this commutator, we once again use the anticommutation relations for the field operators (in this case we 
first differentiate them twice, with respect to the spatial coordinates $x^i$ and $y^i$). The final result is
\begin{eqnarray}
&& \hspace{-1.0cm} \left[\Phi_{\rm GLW}^m(t), \Phi_{\rm GLW}^n(t) \right] 
= \nonumber  \\
&& \frac{2i}{m^2} \varepsilon^{mnj} \int\dx \left( \partial_x^j \psixdag \right) 
\Sigma^l \partial_x^l \psix.
\label{eq:comGLW}
\end{eqnarray}
We show in Appendix~\ref{sec:comGLW} that for the spin polarized systems the expectation value of the operator $C^j$ defined by the space integral in the last line of \rf{eq:comGLW} usually does not vanish. It can be made equal to zero only if very special conditions are fulfilled by the spin distributions of particles. Conseqently, there is no general mechanism that makes the commutator \rfn{eq:comGLW} vanish and the GLW spin operators consistent with the SO(3) algebra.

\section{Conclusions}

\medskip
We conclude by making the following list of observations:

\medskip \noindent
i) The pseudogauge transformations split the total angular momentum into the orbital and spin parts in different ways. In general, the total spin operators obtained in this way may not fulfill the angular momentum algebra SO(3). Out of the three popular pseudogauges considered in this work, only the canonical one fulfills this condition. This makes the canonical pseudogauge especially suitable for the treatment of the spin degrees of freedom. 

\medskip \noindent
ii) Although only the canonical form fulfills the angular momentum algebra for the spin operators, it does not mean that the other pseudogauges are ``wrong''. At the level of the field equations, all pseudogauges are equivalent. The canonical pseudogauge seems to be the most appropriate to describe the spin degrees of freedom whenever a comparison between theory and experiment is made. On the other hand, the other pseudogauges may be favored for other features. For example, they may be more convenient to determine the system's dynamics. In particular, the pseudogauges with the spin tensor conserved offer a possibility of making a direct link to thermodynamic and hydrodynamic approaches~\cite{Florkowski:2018fap}.

\medskip \noindent
iii) Our findings shed new light on the results encountered in recent papers~\cite{Li:2020eon,Buzzegoli:2021wlg,Das:2021aar,Weickgenannt:2022jes}. They are quite difficult to interpret as they provide evidence for both dependence and independence of studied quantities with respect to pseudogauge transformations. A pseudogauge dependence is typically found in quantum calculations~\cite{Buzzegoli:2021wlg,Das:2021aar} while pseudogauge independence or equivalence is commonly found in classical approaches. 
Clearly, a quantum mechanical treatment of spin observables within a pseudogauge that breaks \rf{eq:SO3} is fundamentally inconsistent. It is also imaginable that quantum calculations obtained within two pseudogauges that differ by the fulfillment of \rf{eq:SO3} may lead to different results~\cite{Deriglazov:2016mhk, Wakamatsu:2019ain, Deriglazov:2020ddm}. On the other hand, any classical or semiclassical approaches to spin, where the operator character of the spin observables is neglected, may be found to be fully equivalent. 

\medskip \noindent
iv) In this work, we have not addressed the differences between the canonical and Belinfante pseudogauges. Some of the differences between the results obtained with these two pseudogauges, which have been recently reported in the literature~\cite{Buzzegoli:2021wlg,Becattini:2011ev,Becattini:2018duy}, may be attributed to a reduced description of systems in the Belinfante case as compared to the canonical one. Clearly, this aspect of pseudogauge dependence requires deeper analysis.

\medskip \noindent
v) Other changes of the energy-momentum tensor are considered in the literature, which do not have the form of the pseudogauge transformation given in Eq.~\rfn{eq:PG}. They involve adding a term of the form $\partial_\lambda A^{\lambda \mu \nu}$, with $A^{\lambda \mu \nu} = - A^{\mu \lambda \nu}$ to the energy momentum tensor $T^{\mu\nu}$~\cite{Coleman:2018mew}. It has been recently shown in~Ref.~\cite{Daher:2022xon}, that such a change is necessary to find a relation between the canonical formulation and the so-called phenomenological versions of $T^{\mu\nu}$ and $S^{\lambda, \mu\nu}$.\footnote{In the phenomenological formulation, the spin tensor has the form $S^{\lambda, \mu\nu} = u^\lambda S^{\mu\nu}$~\cite{Weyssenhoff:1947iua,Florkowski:2017ruc}. Although it is sometimes called the canonical version, it does not satisfy the standard condition of the canonical spin tensor that it is totally antisymmetric in all the indices.}
In Ref.~\cite{Fukushima:2020ucl} a connection between the phenomenological and Belinfante versions of the leading-order in gradients spin hydrodynamics was analyzed. The results of Refs.~\cite{Daher:2022xon} and \cite{Fukushima:2020ucl} indicate that an equivalence between the phenomenological, canonical, and Belinfante spin hydrodynamics can be achieved by appropriate redefinitions of the non-equilibrium entropy currents.

\medskip \noindent
vi) Different forms of the spin tensor are discussed in the context of QCD, where a discussion is held on how to split the angular momentum of quarks and gluons into an orbital and a spin part. This topic is broadly discussed in Ref.~\cite{Leader:2013jra}. Our results are supplementary to the QCD studies, where the central issue remains the possibility of a gauge invariant splitting between $\textbf{L}$ and $\textbf{S}$ (realized only with the help of non-local field operators). So far, the pseudogauges discussed in the present work have not been addressed in direct QCD applications.

\medskip\noindent
vii) A natural question can be asked if experimental procedures used to determine the spin polarization of particles can indicate themselves which pseudogauge is the most appropriate to use. In the context of the spin proton puzzle, it has been argued in Ref.~\cite{Leader:2013jra} that the measurability requirement does not solve any of such ambiguities and a particular decomposition of the total angular momentum into the orbital and spin parts is essentially a matter of taste and convenience. In our case, where we analyze a gas of relativistic particles, one considers a measurement of the spin polarization of particles with a given four-momentum $p^\mu = (E, {\pv})$. For theoretical description of such processes one needs to consider the phase space densities of the spin tensor, namely, the quantities $E dS^{\lambda,\mu\nu}(x,p)/d^3p$ which can be obtained by a semiclassical expansion of the Wigner function. It has been shown in Ref.~\cite{Florkowski:2017dyn} that although $S^{\lambda,\mu\nu}(x,p)$ depends on the pseudogauge used to define the spin tensor, the expression for the measured spin polarization  is independent of the pseudogauge (the definition of the Pauli-Luba\'nski vector selects only one term that is common to different pseudogauges). Hence, the case studied in heavy-ion physics is quite similar to that encountered in the QCD proton-spin case -- the measurability argument does not seem to favor any of the pseudogauges and the use of the canonical version is favored by general arguments (locality in the QCD case, good transformation properties in our case). One should, however, keep in mind that the discussed issues are still open and new ideas may appear addressing direct measurements of the operators $S^k(t)$, which would shed a new light on the problem of the pseudogauge selection. 

\bigskip
{\bf Acknowledgements:} We thank David Wagner for stimulating discussions. W.F. and R.R. acknowledge the kind hospitality of the National Institute of Science Education and Research (NISER), Jatni, India, where this analysis was initiated. This work was supported in part by the Polish National Science Center Grant Nos. 2018/30/E/ST2/00432 and 2020/39/D/ST2/02054. A.J. was supported in part by the DST-INSPIRE faculty award under Grant No. DST/INSPIRE/ 04/2017/000038.

\appendix
\section{Commutator of the GLW superpotentials}
\label{sec:comGLW}

\bigskip
To analyze equation \rfn{eq:comGLW} we consider an expansion of the Dirac field operator in terms of creation and annihilation for the helicity states. Following closely the notation introduced in Ref.~\cite{Greiner:1996zu} we write
\begin{eqnarray}
\psi(x)=\sum_{s}\int \!\!dP \left(
b_\pv(s) u_\pv(s)e^{-ip\cdot x} 
+ c^{\dagger}_\pv(s)v_\pv(s)e^{ip\cdot x}\right),
\end{eqnarray}
where $p^\mu = (p^0,\pv)$ with the on-mass-shell energy $p^0 = E_p\!=\!\sqrt{m^2+\pv^2}$, 
and $dP$ is the momentum integration measure
\begin{equation}
 dP = \frac{d^3p}{(2\pi)^{3/2}} \sqrt{\frac{m}{E_p}}.   
\end{equation}
The sum over $s$ includes two helicity states, i.e., the states with a definite spin polarization along the three-momentum vector $\pv$ in the laboratory frame. Following Ref.~\cite{Greiner:1996zu} these states are denoted as $+s$ and $-s$. The helicity bispinors $u_\pv(\pm s)$ and $v_\pv(\pm s)$ satisfy the eigenvalue equations
\begin{eqnarray}
\hat{\sigma}\,u_\pv(\pm s)=\pm\,u_\pv(\pm s), \quad
\hat{\sigma}\,v_\pv(\pm s)=\mp\,v_\pv(\pm s),
\label{eq:ident1}
\end{eqnarray}
where 
\begin{equation}
\hat{\sigma}=\Sigma\cdot\frac{\pv}{|\pv|}
\end{equation}
is the helicity operator. Other useful normalization conditions include
\begin{eqnarray}
&&u^{\dagger}_\pv(s^{\prime})u_\pv(s)
=v^{\dagger}_\pv(s^{\prime})v_\pv(s)
=\frac{E_p}{m}\delta_{s\,s^{\prime}}, \nonumber \\
&& u^{\dagger}_{-\pv}(s^{\prime})v_\pv(s)
=v^{\dagger}_{-\pv}(s^{\prime})u_\pv(s)=0.
\label{eq:ident2}
\end{eqnarray}
The space derivative of the field operator appearing in Eq.~\rfn{eq:comGLW} can be written as
\begin{eqnarray}
\partial_x^j \psi(x) &=& \!\!\! i \sum_{s}\int \!\!dP p^j \left[
b_\pv(s) u_\pv(s)e^{-ip\cdot x} 
\vphantom{c^{\dagger}_\pv(s)} \right. \nonumber \\
&& \left. \,\,\,
- c^{\dagger}_\pv(s)v_\pv(s)e^{ip\cdot x}\right].
\label{eq:psider}
\end{eqnarray}

\medskip
The expressions given above can be directly used to calculate the operator $C^j$ defined by the space integral on the right-hand side of \rf{eq:comGLW}. Space integration of the product of the operators of the form \rfn{eq:psider} with the operator $\Sigma$ inserted in between them gives
\begin{eqnarray}
C^{j} &=& \int d^{3}p \, p^{j}|\pv|
    \left[ b^{\dagger}_\pv(+s) b_\pv(+s)
    -b^{\dagger}_\pv(-s)b_\pv(-s) \right.
    \nonumber \\
&&  \left.  -c_\pv(+s) c^{\dagger}_\pv(+s)
+c_\pv(-s) c^{\dagger}_\pv(-s) \right].
\label{eq:Cj1}
\end{eqnarray}
Here we have used Eqs.~\rfn{eq:ident1} and \rfn{eq:ident2}.

The standard procedure at this point is to subtract an infinite constant from \rf{eq:Cj1}, which is equivalent to normal ordering that gives~\footnote{One can also argue that this constant is in fact zero as it follows from the integral of the form $\int d^{3}p \, p^{j}|\pv|$.}
\begin{eqnarray}
:C^{j}: &=& \int d^{3}p \, p^{j}|\pv|
    \left[ b^{\dagger}_\pv(+s) b_\pv(+s)
    -b^{\dagger}_\pv(-s)b_\pv(-s) \right.
    \nonumber \\
&&  \left.  + c^{\dagger}_\pv(+s) c_\pv(+s) 
-  c^{\dagger}_\pv(-s) c_\pv(-s) \right].
\label{eq:Cj2}
\end{eqnarray}
The operators $b^{\dagger}_\pv(\pm s) b_\pv(\pm s)$ and $c^{\dagger}_\pv(\pm s) c_\pv(\pm s) $ play role of occupation number operators of particles with helicities $\pm 1$. 

In general, the expectation value of the operator $:C^{j}:$ is not zero. There are, however, several cases where it vanishes. This is so, for example, in the case where the occupation numbers of particles with positive and negative helicities are equal. Then, the terms with opposite signs on the right-hand side of \rf{eq:Cj2} cancel each other. This means, however, that the system as a whole does not have any net value of the spin polarization, hence, it is rather not interesting for the spin studies. Similar situation takes place where the occupation numbers are even functions of momentum components. In this case, due to the appearance of $p^j$ in the integrand, the momentum integration in \rf{eq:Cj2} gives zero. In this case we deal again with the system that is not spin-polarized as a whole. For spin-polarized systems which do not exhibit any symmetries, one expects that the expectation value of $:C^{j}:$ does not vanish.

 
\printcredits

\bibliographystyle{model6-num-names}

\bibliography{cas-refs}

\begin{thebibliography}{29}
\providecommand{\natexlab}[1]{#1}
\providecommand{\url}[1]{\texttt{#1}}
\providecommand{\href}[2]{#2}
\providecommand{\path}[1]{#1}
\providecommand{\DOIprefix}{doi:}
\providecommand{\ArXivprefix}{arXiv:}
\providecommand{\URLprefix}{URL: }
\providecommand{\Pubmedprefix}{pmid:}
\providecommand{\doi}[1]{\href{http://dx.doi.org/#1}{\path{#1}}}
\providecommand{\Pubmed}[1]{\href{pmid:#1}{\path{#1}}}
\providecommand{\BIBand}{and}
\providecommand{\bibinfo}[2]{#2}
\ifx\xfnm\undefined \def\xfnm[#1]{\unskip,\space#1}\fi
\makeatletter\def\@biblabel#1{#1.}\makeatother
\bibitem[{Florkowski et~al.(2019)Florkowski, Kumar and
  Ryblewski}]{Florkowski:2018fap}
\bibinfo{author}{Florkowski\xfnm[ W.]}, \bibinfo{author}{Kumar\xfnm[ A.]},
  \bibinfo{author}{Ryblewski\xfnm[ R.]}.
\newblock \bibinfo{title}{{Relativistic hydrodynamics for spin-polarized
  fluids}}.
\newblock \emph{\bibinfo{journal}{Prog Part Nucl Phys}};
  \bibinfo{year}{2019};\bibinfo{volume}{108}:\bibinfo{pages}{103709}.
\newblock \DOIprefix\doi{10.1016/j.ppnp.2019.07.001};
  \href{http://arxiv.org/abs/1811.04409}{\tt arXiv:1811.04409}.
\bibitem[{Speranza and Weickgenannt(2021)}]{Speranza:2020ilk}
\bibinfo{author}{Speranza\xfnm[ E.]}, \bibinfo{author}{Weickgenannt\xfnm[ N.]}.
\newblock \bibinfo{title}{{Spin tensor and pseudo-gauges: from nuclear
  collisions to gravitational physics}}.
\newblock \emph{\bibinfo{journal}{Eur Phys J A}};
  \bibinfo{year}{2021};\bibinfo{volume}{57}(\bibinfo{number}{5}):\bibinfo{pages}{155}.
\newblock \DOIprefix\doi{10.1140/epja/s10050-021-00455-2};
  \href{http://arxiv.org/abs/2007.00138}{\tt arXiv:2007.00138}.
\bibitem[{Bhadury et~al.(2021)Bhadury, Bhatt, Jaiswal and
  Kumar}]{Bhadury:2021oat}
\bibinfo{author}{Bhadury\xfnm[ S.]}, \bibinfo{author}{Bhatt\xfnm[ J.]},
  \bibinfo{author}{Jaiswal\xfnm[ A.]}, \bibinfo{author}{Kumar\xfnm[ A.]}.
\newblock \bibinfo{title}{{New developments in relativistic fluid dynamics with
  spin}}.
\newblock \emph{\bibinfo{journal}{Eur Phys J ST}};
  \bibinfo{year}{2021};\bibinfo{volume}{230}(\bibinfo{number}{3}):\bibinfo{pages}{655--672}.
\newblock \DOIprefix\doi{10.1140/epjs/s11734-021-00020-4};
  \href{http://arxiv.org/abs/2101.11964}{\tt arXiv:2101.11964}.
\bibitem[{Leader and Lorcé(2014)}]{Leader:2013jra}
\bibinfo{author}{Leader\xfnm[ E.]}, \bibinfo{author}{Lorcé\xfnm[ C.]}.
\newblock \bibinfo{title}{{The angular momentum controversy: What’s it all
  about and does it matter?}}
\newblock \emph{\bibinfo{journal}{Phys Rept}};
  \bibinfo{year}{2014};\bibinfo{volume}{541}(\bibinfo{number}{3}):\bibinfo{pages}{163--248}.
\newblock \DOIprefix\doi{10.1016/j.physrep.2014.02.010};
  \href{http://arxiv.org/abs/1309.4235}{\tt arXiv:1309.4235}.
\bibitem[{Hehl(1976)}]{Hehl:1976vr}
\bibinfo{author}{Hehl\xfnm[ F.W.]}.
\newblock \bibinfo{title}{{On the Energy Tensor of Spinning Massive Matter in
  Classical Field Theory and General Relativity}}.
\newblock \emph{\bibinfo{journal}{Rept Math Phys}};
  \bibinfo{year}{1976};\bibinfo{volume}{9}:\bibinfo{pages}{55--82}.
\newblock \DOIprefix\doi{10.1016/0034-4877(76)90016-1}.
\bibitem[{Belinfante(1939)}]{BELINFANTE1939887}
\bibinfo{author}{Belinfante\xfnm[ F.]}.
\newblock \bibinfo{title}{On the spin angular momentum of mesons}.
\newblock \emph{\bibinfo{journal}{Physica}};
  \bibinfo{year}{1939};\bibinfo{volume}{6}(\bibinfo{number}{7}):\bibinfo{pages}{887--898}.
\newblock \URLprefix
  \url{https://www.sciencedirect.com/science/article/pii/S003189143990090X};
  \DOIprefix\doi{https://doi.org/10.1016/S0031-8914(39)90090-X}.
\bibitem[{Belinfante(1940)}]{Belinfante:1940}
\bibinfo{author}{Belinfante\xfnm[ F.]}.
\newblock \bibinfo{title}{On the current and the density of the electric
  charge, the energy, the linear momentum and the angular momentum of arbitrary
  fields}.
\newblock \emph{\bibinfo{journal}{Physica}};
  \bibinfo{year}{1940};\bibinfo{volume}{7}(\bibinfo{number}{5}):\bibinfo{pages}{449
  -- 474}.
\newblock \URLprefix
  \url{http://www.sciencedirect.com/science/article/pii/S003189144090091X};
  \DOIprefix\doi{https://doi.org/10.1016/S0031-8914(40)90091-X}.
\bibitem[{Rosenfeld(1940)}]{rosenfeld1940tenseur}
\bibinfo{author}{Rosenfeld\xfnm[ L.J.H.C.]}.
\newblock \bibinfo{title}{Sur le tenseur d'impulsion-{\'e}nergie}.
\newblock \bibinfo{publisher}{Palais des acad{\'e}mies}; \bibinfo{year}{1940}.
\bibitem[{Itzykson and Zuber(1980)}]{Itzykson:1980rh}
\bibinfo{author}{Itzykson\xfnm[ C.]}, \bibinfo{author}{Zuber\xfnm[ J.B.]}.
\newblock \bibinfo{title}{{Quantum Field Theory}}.
\newblock International Series In Pure and Applied Physics;
  \bibinfo{address}{New York}: \bibinfo{publisher}{McGraw-Hill};
  \bibinfo{year}{1980}.
\newblock ISBN \bibinfo{isbn}{9780486445687, 0486445682}.
\newblock \URLprefix \url{http://dx.doi.org/10.1063/1.2916419}.
\bibitem[{De~Groot(1980)}]{DeGroot:1980dk}
\bibinfo{author}{De~Groot\xfnm[ S.R.]}.
\newblock \bibinfo{title}{{Relativistic Kinetic Theory. Principles and
  Applications}}.
\newblock \bibinfo{year}{1980}.
\bibitem[{Hilgevoord and Wouthuysen(1963)}]{HILGEVOORD19631}
\bibinfo{author}{Hilgevoord\xfnm[ J.]}, \bibinfo{author}{Wouthuysen\xfnm[ S.]}.
\newblock \bibinfo{title}{On the spin angular momentum of the dirac particle}.
\newblock \emph{\bibinfo{journal}{Nuclear Physics}};
  \bibinfo{year}{1963};\bibinfo{volume}{40}:\bibinfo{pages}{1--12}.
\newblock \URLprefix
  \url{https://www.sciencedirect.com/science/article/pii/0029558263902463};
  \DOIprefix\doi{https://doi.org/10.1016/0029-5582(63)90246-3}.
\bibitem[{Hilgevoord and {De Kerf}(1965)}]{HILGEVOORD19651002}
\bibinfo{author}{Hilgevoord\xfnm[ J.]}, \bibinfo{author}{{De Kerf}\xfnm[ E.]}.
\newblock \bibinfo{title}{The covariant definition of spin in relativistic
  quantum field theory}.
\newblock \emph{\bibinfo{journal}{Physica}};
  \bibinfo{year}{1965};\bibinfo{volume}{31}(\bibinfo{number}{7}):\bibinfo{pages}{1002--1016}.
\newblock \URLprefix
  \url{https://www.sciencedirect.com/science/article/pii/0031891465901412};
  \DOIprefix\doi{https://doi.org/10.1016/0031-8914(65)90141-2}.
\bibitem[{Leader and Lorcé(2019)}]{LEADER201923}
\bibinfo{author}{Leader\xfnm[ E.]}, \bibinfo{author}{Lorcé\xfnm[ C.]}.
\newblock \bibinfo{title}{Corrigendum to “the angular momentum controversy:
  What’s it all about and does it matter?” [phys. rep. 541(3) (2014)
  163–248]}.
\newblock \emph{\bibinfo{journal}{Physics Reports}};
  \bibinfo{year}{2019};\bibinfo{volume}{802}:\bibinfo{pages}{23--24}.
\newblock \URLprefix
  \url{https://www.sciencedirect.com/science/article/pii/S0370157319300328};
  \DOIprefix\doi{https://doi.org/10.1016/j.physrep.2019.01.006}.
\bibitem[{Coleman(2018)}]{Coleman:2018mew}
\bibinfo{author}{Coleman\xfnm[ S.]}.
\newblock \bibinfo{title}{{Lectures of Sidney Coleman on Quantum Field
  Theory}}.
\newblock \bibinfo{address}{Hackensack}: \bibinfo{publisher}{WSP};
  \bibinfo{year}{2018}.
\newblock ISBN \bibinfo{isbn}{978-981-4632-53-9, 978-981-4635-50-9}.
\newblock \DOIprefix\doi{10.1142/9371}.
\bibitem[{Li et~al.(2021)Li, Stephanov and Yee}]{Li:2020eon}
\bibinfo{author}{Li\xfnm[ S.]}, \bibinfo{author}{Stephanov\xfnm[ M.A.]},
  \bibinfo{author}{Yee\xfnm[ H.U.]}.
\newblock \bibinfo{title}{{Nondissipative Second-Order Transport, Spin, and
  Pseudogauge Transformations in Hydrodynamics}}.
\newblock \emph{\bibinfo{journal}{Phys Rev Lett}};
  \bibinfo{year}{2021};\bibinfo{volume}{127}(\bibinfo{number}{8}):\bibinfo{pages}{082302}.
\newblock \DOIprefix\doi{10.1103/PhysRevLett.127.082302};
  \href{http://arxiv.org/abs/2011.12318}{\tt arXiv:2011.12318}.
\bibitem[{Buzzegoli(2022)}]{Buzzegoli:2021wlg}
\bibinfo{author}{Buzzegoli\xfnm[ M.]}.
\newblock \bibinfo{title}{{Pseudogauge dependence of the spin polarization and
  of the axial vortical effect}}.
\newblock \emph{\bibinfo{journal}{Phys Rev C}};
  \bibinfo{year}{2022};\bibinfo{volume}{105}(\bibinfo{number}{4}):\bibinfo{pages}{044907}.
\newblock \DOIprefix\doi{10.1103/PhysRevC.105.044907};
  \href{http://arxiv.org/abs/2109.12084}{\tt arXiv:2109.12084}.
\bibitem[{Das et~al.(2021)Das, Florkowski, Ryblewski and Singh}]{Das:2021aar}
\bibinfo{author}{Das\xfnm[ A.]}, \bibinfo{author}{Florkowski\xfnm[ W.]},
  \bibinfo{author}{Ryblewski\xfnm[ R.]}, \bibinfo{author}{Singh\xfnm[ R.]}.
\newblock \bibinfo{title}{{Pseudogauge dependence of quantum fluctuations of
  the energy in a hot relativistic gas of fermions}}.
\newblock \emph{\bibinfo{journal}{Phys Rev D}};
  \bibinfo{year}{2021};\bibinfo{volume}{103}(\bibinfo{number}{9}):\bibinfo{pages}{L091502}.
\newblock \DOIprefix\doi{10.1103/PhysRevD.103.L091502};
  \href{http://arxiv.org/abs/2103.01013}{\tt arXiv:2103.01013}.
\bibitem[{Weickgenannt et~al.(2022)Weickgenannt, Wagner and
  Speranza}]{Weickgenannt:2022jes}
\bibinfo{author}{Weickgenannt\xfnm[ N.]}, \bibinfo{author}{Wagner\xfnm[ D.]},
  \bibinfo{author}{Speranza\xfnm[ E.]}.
\newblock \bibinfo{title}{{Pseudogauges and relativistic spin hydrodynamics for
  interacting Dirac and Proca fields}}.
\newblock \emph{\bibinfo{journal}{Phys Rev D}};
  \bibinfo{year}{2022};\bibinfo{volume}{105}(\bibinfo{number}{11}):\bibinfo{pages}{116026}.
\newblock \DOIprefix\doi{10.1103/PhysRevD.105.116026};
  \href{http://arxiv.org/abs/2204.01797}{\tt arXiv:2204.01797}.
\bibitem[{Deriglazov and Pupasov-Maksimov(2016)}]{Deriglazov:2016mhk}
\bibinfo{author}{Deriglazov\xfnm[ A.A.]},
  \bibinfo{author}{Pupasov-Maksimov\xfnm[ A.M.]}.
\newblock \bibinfo{title}{{Relativistic corrections to the algebra of position
  variables and spin-orbital interaction}}.
\newblock \emph{\bibinfo{journal}{Phys Lett B}};
  \bibinfo{year}{2016};\bibinfo{volume}{761}:\bibinfo{pages}{207--212}.
\newblock \DOIprefix\doi{10.1016/j.physletb.2016.08.034};
  \href{http://arxiv.org/abs/1609.00043}{\tt arXiv:1609.00043}.
\bibitem[{Wakamatsu(2019)}]{Wakamatsu:2019ain}
\bibinfo{author}{Wakamatsu\xfnm[ M.]}.
\newblock \bibinfo{title}{{A still unsettled issue in the nucleon spin
  decomposition problem : On the role of surface terms and gluon topology}}.
\newblock \emph{\bibinfo{journal}{Eur Phys J A}};
  \bibinfo{year}{2019};\bibinfo{volume}{55}(\bibinfo{number}{7}):\bibinfo{pages}{123}.
\newblock \DOIprefix\doi{10.1140/epja/i2019-12800-9};
  \href{http://arxiv.org/abs/1905.03412}{\tt arXiv:1905.03412}.
\bibitem[{Deriglazov(2020)}]{Deriglazov:2020ddm}
\bibinfo{author}{Deriglazov\xfnm[ A.A.]}.
\newblock \bibinfo{title}{{Nonminimal Spin-Field Interaction of the Classical
  Electron and Quantization of Spin}}.
\newblock \emph{\bibinfo{journal}{Phys Part Nucl Lett}};
  \bibinfo{year}{2020};\bibinfo{volume}{17}(\bibinfo{number}{5}):\bibinfo{pages}{738--743}.
\newblock \DOIprefix\doi{10.1134/S1547477120050131};
  \href{http://arxiv.org/abs/2001.01294}{\tt arXiv:2001.01294}.
\bibitem[{Becattini and Tinti(2011)}]{Becattini:2011ev}
\bibinfo{author}{Becattini\xfnm[ F.]}, \bibinfo{author}{Tinti\xfnm[ L.]}.
\newblock \bibinfo{title}{{Thermodynamical inequivalence of quantum
  stress-energy and spin tensors}}.
\newblock \emph{\bibinfo{journal}{Phys Rev D}};
  \bibinfo{year}{2011};\bibinfo{volume}{84}:\bibinfo{pages}{025013}.
\newblock \DOIprefix\doi{10.1103/PhysRevD.84.025013};
  \href{http://arxiv.org/abs/1101.5251}{\tt arXiv:1101.5251}.
\bibitem[{Becattini et~al.(2019)Becattini, Florkowski and
  Speranza}]{Becattini:2018duy}
\bibinfo{author}{Becattini\xfnm[ F.]}, \bibinfo{author}{Florkowski\xfnm[ W.]},
  \bibinfo{author}{Speranza\xfnm[ E.]}.
\newblock \bibinfo{title}{{Spin tensor and its role in non-equilibrium
  thermodynamics}}.
\newblock \emph{\bibinfo{journal}{Phys Lett}};
  \bibinfo{year}{2019};\bibinfo{volume}{B789}:\bibinfo{pages}{419--425}.
\newblock \DOIprefix\doi{10.1016/j.physletb.2018.12.016};
  \href{http://arxiv.org/abs/1807.10994}{\tt arXiv:1807.10994}.
\bibitem[{Daher et~al.(2022)Daher, Das, Florkowski and
  Ryblewski}]{Daher:2022xon}
\bibinfo{author}{Daher\xfnm[ A.]}, \bibinfo{author}{Das\xfnm[ A.]},
  \bibinfo{author}{Florkowski\xfnm[ W.]}, \bibinfo{author}{Ryblewski\xfnm[
  R.]}.
\newblock \bibinfo{title}{{Canonical and phenomenological formulations of spin
  hydrodynamics}}.
\newblock \bibinfo{year}{2022};\href{http://arxiv.org/abs/2202.12609}{\tt
  arXiv:2202.12609}.
\bibitem[{Weyssenhoff and Raabe(1947)}]{Weyssenhoff:1947iua}
\bibinfo{author}{Weyssenhoff\xfnm[ J.]}, \bibinfo{author}{Raabe\xfnm[ A.]}.
\newblock \bibinfo{title}{{Relativistic dynamics of spin-fluids and
  spin-particles}}.
\newblock \emph{\bibinfo{journal}{Acta Phys Polon}};
  \bibinfo{year}{1947};\bibinfo{volume}{9}:\bibinfo{pages}{7--18}.
\bibitem[{Florkowski et~al.(2018{\natexlab{a}})Florkowski, Friman, Jaiswal and
  Speranza}]{Florkowski:2017ruc}
\bibinfo{author}{Florkowski\xfnm[ W.]}, \bibinfo{author}{Friman\xfnm[ B.]},
  \bibinfo{author}{Jaiswal\xfnm[ A.]}, \bibinfo{author}{Speranza\xfnm[ E.]}.
\newblock \bibinfo{title}{{Relativistic fluid dynamics with spin}}.
\newblock \emph{\bibinfo{journal}{Phys Rev}};
  \bibinfo{year}{2018}{\natexlab{a}};\bibinfo{volume}{C97}(\bibinfo{number}{4}):\bibinfo{pages}{041901}.
\newblock \DOIprefix\doi{10.1103/PhysRevC.97.041901};
  \href{http://arxiv.org/abs/1705.00587}{\tt arXiv:1705.00587}.
\bibitem[{Fukushima and Pu(2021)}]{Fukushima:2020ucl}
\bibinfo{author}{Fukushima\xfnm[ K.]}, \bibinfo{author}{Pu\xfnm[ S.]}.
\newblock \bibinfo{title}{{Spin hydrodynamics and symmetric energy-momentum
  tensors \textendash{} A current induced by the spin vorticity
  \textendash{}}}.
\newblock \emph{\bibinfo{journal}{Phys Lett B}};
  \bibinfo{year}{2021};\bibinfo{volume}{817}:\bibinfo{pages}{136346}.
\newblock \DOIprefix\doi{10.1016/j.physletb.2021.136346};
  \href{http://arxiv.org/abs/2010.01608}{\tt arXiv:2010.01608}.
\bibitem[{Florkowski et~al.(2018{\natexlab{b}})Florkowski, Friman, Jaiswal,
  Ryblewski and Speranza}]{Florkowski:2017dyn}
\bibinfo{author}{Florkowski\xfnm[ W.]}, \bibinfo{author}{Friman\xfnm[ B.]},
  \bibinfo{author}{Jaiswal\xfnm[ A.]}, \bibinfo{author}{Ryblewski\xfnm[ R.]},
  \bibinfo{author}{Speranza\xfnm[ E.]}.
\newblock \bibinfo{title}{{Spin-dependent distribution functions for
  relativistic hydrodynamics of spin-1/2 particles}}.
\newblock \emph{\bibinfo{journal}{Phys Rev}};
  \bibinfo{year}{2018}{\natexlab{b}};\bibinfo{volume}{D97}(\bibinfo{number}{11}):\bibinfo{pages}{116017}.
\newblock \DOIprefix\doi{10.1103/PhysRevD.97.116017};
  \href{http://arxiv.org/abs/1712.07676}{\tt arXiv:1712.07676}.
\bibitem[{Greiner and Reinhardt(1996)}]{Greiner:1996zu}
\bibinfo{author}{Greiner\xfnm[ W.]}, \bibinfo{author}{Reinhardt\xfnm[ J.]}.
\newblock \bibinfo{title}{{Field quantization}}.
\newblock \bibinfo{year}{1996}.

\end{thebibliography}
\end{document}